\documentclass[iop]{emulateapj}

\usepackage{natbib}

\usepackage{epstopdf}
\usepackage{amsmath}

\newcommand{\mc}{\multicolumn{2}{c}}

\newcommand{\Ha}{\ensuremath{\mathrm{H} \alpha}}

\newcommand{\vlsr}{\ensuremath{v_{\mathrm{LSR}}}}
\newcommand{\vgsr}{\ensuremath{v_{\mathrm{GSR}}}}

\newcommand{\sii}{\ensuremath{\textrm{[\ion{S}{2}]}}}
\newcommand{\nii}{\ensuremath{\textrm{[\ion{N}{2}]}}}
\newcommand{\oiii}{\ensuremath{\textrm{[\ion{O}{3}]}}}

\newcommand{\EM}{\ensuremath{\mathrm{EM}}}

\newcommand{\RM}{\ensuremath{\mathrm{RM}}}
\newcommand{\RMHVC}{\ensuremath{\RM_{\mathrm{HVC}}}}

\newcommand{\R}{\ensuremath{\, \mathrm{R}}}
\newcommand{\Hp}{\ensuremath{\mathrm{H}^+}}

\newcommand{\RMt}{\ensuremath{\langle \RMHVC \rangle}}

\newcommand{\cm}{\ensuremath{\textrm{ cm}}}

\newcommand{\radmsq}{\ensuremath{\textrm{ rad m}^{-2}}}
\newcommand{\cmsix}{\ensuremath{\cm^{-6}}}
\newcommand{\kms}{\ensuremath{\textrm{ km s}^{-1}}}
\newcommand{\kpc}{\ensuremath{\, \mathrm{kpc}}}
\newcommand{\pc}{\ensuremath{\textrm{ pc}}}
\newcommand{\K}{\ensuremath{\textrm{ K}}}
\newcommand{\uG}{\ensuremath{\, \mu \mathrm{G}}}

\renewcommand{\vec}{\mathbf}

\begin{document}

\date{\today} 

\submitted{Accepted for publication in the Astrophysical Journal}
\title{Magnetized gas in the Smith High Velocity Cloud}
\author{Alex S. Hill\altaffilmark{1}, S. A. Mao\altaffilmark{2,3}, Robert A. Benjamin\altaffilmark{4}, Felix J. Lockman\altaffilmark{5}, \& Naomi M. McClure-Griffiths\altaffilmark{1}}
\altaffiltext{1}{CSIRO Astronomy \& Space Science, Marsfield, NSW, Australia; alex.hill@csiro.au, naomi.mcclure-griffiths@csiro.au}
\altaffiltext{2}{Department of Astronomy, University of Wisconsin-Madison, Madison, WI, USA; mao@astro.wisc.edu}
\altaffiltext{3}{Jansky Fellow, National Radio Astronomy Observatory, Socorro, NM, USA}
\altaffiltext{4}{Department of Physics, University of Wisconsin-Whitewater, Whitewater, WI, USA; benjamir@uww.edu}
\altaffiltext{5}{National Radio Astronomy Observatory, Green Bank, WV, USA; jlockman@nrao.edu}

\begin{abstract}
We report the first detection of magnetic fields associated with the Smith High Velocity Cloud. We use a catalog of Faraday rotation measures towards extragalactic radio sources behind the Smith Cloud, new \ion{H}{1} observations from the Green Bank Telescope, and a  spectroscopic map of \Ha\ from the Wisconsin H-Alpha Mapper Northern Sky Survey. There are enhancements in rotation measure of $\approx 100 \radmsq$ which are generally well correlated with decelerated \Ha\ emission. We estimate a lower limit on the line-of-sight component of the field of $\approx 8 \uG$ along a decelerated filament; this is a lower limit due to our assumptions about the geometry. No \RM\ excess is evident in sightlines dominated by \ion{H}{1} or \Ha\ at the velocity of the Smith Cloud. The smooth \Ha\ morphology of the emission at the Smith Cloud velocity suggests photoionization by the Galactic ionizing radiation field as the dominant ionization mechanism, while the filamentary morphology and high ($\approx 1$ Rayleigh) \Ha\ intensity of the lower-velocity magnetized ionized gas suggests an ionization process associated with shocks due to interaction with the Galactic interstellar medium. The presence of the magnetic field may contribute to the survival of high velocity clouds like the Smith Cloud as they move from the Galactic halo to the disk. We expect these data to provide a test for magnetohydrodynamic simulations of infalling gas. 
\end{abstract}

\maketitle

\section{Introduction}

Infall of gas drives the evolution of galaxies \citep[e.g.][]{Wakker:1997ha,Blitz:1999hv,Putman:2012bs} and may be necessary to explain the observed mass-metallicity relationship \citep{Tremonti:2004ed}. High velocity clouds (HVCs) likely trace this inflow. HVCs are gas observed at velocities inconsistent with Galactic rotation, defined variously in terms of their local standard of rest (LSR: $|\vlsr| > 90 \kms$) or deviation ($|v_{\mathrm{dev}}| > 50 \kms$) velocity \citep{Wakker:1991ut,Wakker:1997ha,Moss:2013vy}. 
Though primarily studied in \ion{H}{1} \citep{Wakker:1997ha}, HVCs have a substantial or dominant ionized component, some at $\sim 10^4 \K$ \citep[e.g.][]{Tufte:1998cb,Tufte:2002du,BlandHawthorn:2007kc,Hill:2009gx,Barger:2012jz} and some more highly ionized \citep[e.g.][]{Sembach:2000ib,Sembach:2003dr,Fox:2006kg,Shull:2009gs,Lehner:2011ic}.

The Smith Cloud is perhaps the best HVC for tracing the interaction between the Galactic halo and interstellar medium (ISM). It has a cometary morphology, with a head near $(l, b) = (39\arcdeg, -13\arcdeg)$ and a tail extending well past $(45\arcdeg, -20\arcdeg)$. From three independent measurements, the distance to its head is $12.4 \pm 1.3 \kpc$ \citep[][hereafter L08]{Wakker:2008fo,Putman:2003ba,Lockman:2008ha}, placing it $3 \kpc$ below the Galactic plane and moving upwards towards the disk. L08 also derived all three components of the space velocity of the cloud. The \ion{H}{1} mass of gas mapped by L08 is $10^6 M_\odot$, though there is considerable emission outside the area they observed as well. While the origin of the cloud is unknown, it has been the subject of much speculation including an association with the Sgr dwarf \citep{BlandHawthorn:1998kb} and a jet from the molecular ring \citep{Sofue:2004wq}. \citet{Nichols:2009im} suggested that an associated dark matter halo could confine the cloud. Its low nitrogen abundance suggests an extragalactic origin \citep[but see discussion in Section~\ref{sec:shocks} below]{Hill:2009gx}, while its prograde orbit inclined at only $30\arcdeg$ suggests knowledge of the Galactic gravitational potential \citep[][L08]{Smith:1963}. Its calculated orbit suggests that it survived a passage through the Galactic disk at a Galactocentric radius of $\approx 13 \kpc$ (L08).

Hydrodynamical simulations suggest that HVCs like the Smith Cloud should lose most of their neutral hydrogen content after traveling $\lesssim 10 \kpc$ through the Galactic halo and that clouds need to be denser than the ambient medium by a factor of $\sim 10-20$ to survive infall \citep{Keres:2009iu,Heitsch:2009do,Joung:2012ea}. If HVCs are disrupted on these scales, much of the accreting gas likely feeds the warm ionized medium \citep{Benjamin:1997hm,BlandHawthorn:2009js,Kwak:2011hy,Henley:2012hy}. However, although the Smith Cloud likely has an ionized envelope as massive as the neutral cloud \citep{Hill:2009gx} and the \ion{H}{1} is disrupted, $\gtrsim 10^6 M_\odot$ remains concentrated in a coherent \ion{H}{1} structure.


The few simulations of HVCs to date which have explored the effects of magnetic fields in two \citep{Konz:2002cy} or three \citep{Santillan:2004wi} dimensions (2D or 3D) support analytic expectations that magnetic fields of at least a few \uG\ can stabilize a cloud against disruption. In 2D simulations, \citet{Santillan:1999dk} showed that a magnetized disk can prevent an unmagmetized HVC from penetrating into the disk due to increased magnetic tension from compressed field lines, although the interaction drives waves on both sides of the disk. The orientation of the field both within the cloud and in the ambient ISM substantially affects its evolution \citep{Kwak:2009gm}, so observations of the magnetic field geometry within HVCs are a necessary constraint on accretion models.

Measuring magnetic fields in interstellar gas is difficult. Attempts to measure Zeeman splitting of the \ion{H}{1} 21~cm line, the best direct measure of magnetic fields in neutral hydrogen, have not been successful for the Smith Cloud due to potential structure in \ion{H}{1} within the beam. Faraday rotation measures the integrated line of sight magnetic field in ionized gas weighted by the electron density. The measured rotation in the linear polarization angle of a background radio source at wavelength $\lambda$ is $\Delta \theta = \RM \lambda^2$, where the rotation measure is
\begin{equation}
\RM = 0.81 \int_\mathrm{source}^\mathrm{observer} n_e(s) \vec{B}(s) \cdot d\vec{s} \radmsq.
\end{equation}
Here $n_e(s)$ is the electron density as a function of position along the line of sight (in $\mathrm{cm}^3$), $\vec{B}(s)$ is the magnetic field (in\uG), and $\vec{s}$ is the line-of-sight vector (in pc) from the source to the observer; thus, a positive RM indicates a field pointing towards the observer.

A recent all-northern-sky catalog of RMs towards extragalactic points sources \citep{Taylor:2009hv} has allowed many studies of the interstellar magnetic field. \citet{McClureGriffiths:2010fc} used these data to measure a line-of-sight magnetic field of $B_{||} \ge 6 \uG$ in an HVC in the Leading Arm of the Magellanic System. Here, we apply the technique of \citet{McClureGriffiths:2010fc} to the Smith Cloud to measure the magnetic field in a cloud interacting with the Galactic disk. In Section~\ref{sec:data}, we describe the \ion{H}{1}, \Ha\, and Faraday rotation data we use. We describe the spatial and kinematic structure of the Smith Cloud in Section~\ref{sec:overview} and argue that essentially all of the emission studied here is physically associated with the cloud in Section~\ref{sec:interaction}. We then estimate electron densities and magnetic field strengths in Section~\ref{sec:B}. We discuss our results in Section~\ref{sec:discussion} and summarize the paper in Section~\ref{sec:summary}.

\section{Data} \label{sec:data}


For Faraday rotation, we use the catalog of \citet{Taylor:2009hv}. They derived RMs from the NRAO VLA Sky Survey \citep[NVSS;][]{Condon:1998kn} for $\approx 1$ extragalactic point source per square degree. They quote a typical uncertainty in individual RM measurements of $1-2 \radmsq$ for $|\RM| \lesssim 650 \radmsq$. At larger RMs, a wrapping ambiguity can be of concern because the NVSS polarization data sample only two frequencies. However, such large RMs are unlikely at the Galactic latitude of the Smith Cloud: at $-30\arcdeg < b < -10\arcdeg$, \citet{Taylor:2009hv} list mean RMs of $\approx -20 \radmsq$ with a standard deviation of $< 90 \radmsq$, so $|\RM| = 650 \radmsq$ is $7 \sigma$ from the mean of the distribution. \citet{Mao:2010eg,Mao:2012kq}, using measurements which better sample $\lambda^2$ space, have confirmed that high latitude RMs from \citet{Taylor:2009hv} do not suffer from the wrapping ambiguity.


The \ion{H}{1} data used here comes from a new survey of the Smith Cloud and its environs made with the 100 meter Robert C. Byrd Green Bank Telescope (GBT) of the NRAO.  The spectra cover $700 \kms$ around zero velocity LSR. The velocity resolution is $0.65 \kms$, and the typical rms noise level is $90 \textrm{ mK}$ in a $0.65 \kms$ channel. The angular resolution is $9.1$~arcmin.  Data have been calibrated and corrected for stray radiation as described by \citet{Boothroyd:2011fs}.  The new survey, which will be described in full elsewhere, covers an area more than twice that of the original (L08) survey and detects many regions of extended Smith Cloud emission that were previously unknown.


We use \Ha\ spectroscopic maps from the Wisconsin H-Alpha Mapper (WHAM) Northern Sky Survey \citep[WHAM-NSS;][]{Haffner:2003fe}. These data cover the entire region we consider here with a $3 \sigma$ sensitivity of $\approx 0.15 \textrm{ R}$ and a spectral resolution of $12 \kms$. The $1\arcdeg$ WHAM beam is spaced on a grid with $\approx 1\arcdeg$ between pointing centers, undersampling the image. The map presented here covers $\vlsr \lesssim +70 \kms$. \citet{Hill:2009gx} presented a WHAM \Ha\ map for a subset of this region covering the range $-70 \kms \lesssim \vlsr \lesssim +130 \kms$. We have considered these data in our analysis but do not present them here. An estimated extinction correction in the sightlines and latitudes we consider would add $\approx 40 \%$ to the true \Ha\ intensities at $b \approx -13\arcdeg$ \citep{Hill:2009gx} and less at higher latitudes. Because this extinction correction is highly uncertain and affects derived densities and magnetic field strengths by $\lesssim 18 \%$ (see equations~\ref{eq:ne} and \ref{eq:B} below), we do not perform an extinction correction.

\begin{figure*}[tb]
\plotone{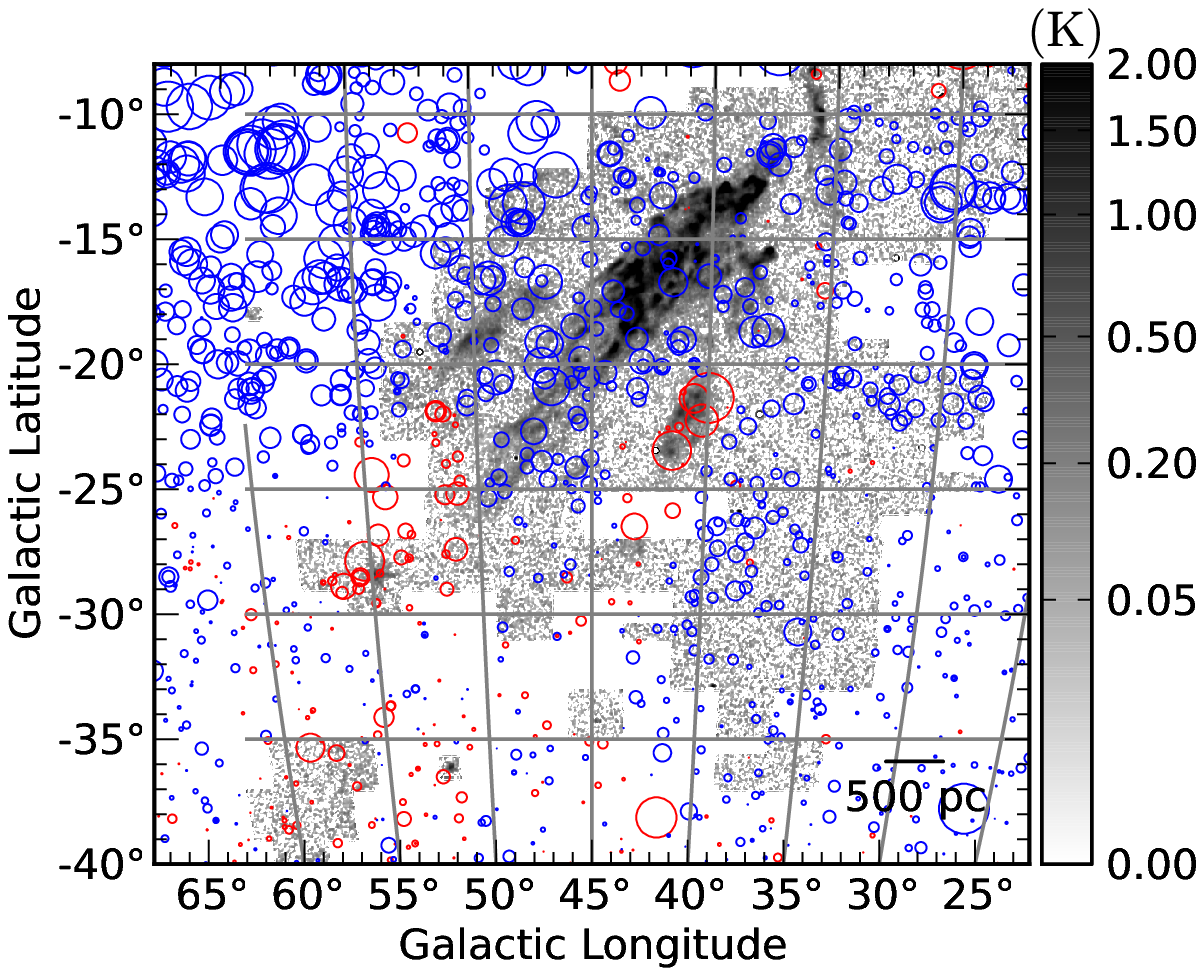}
\caption{\citet{Taylor:2009hv} RMs overlaid on GBT \ion{H}{1} data. The grayscale shows \ion{H}{1} emission in the $\vgsr = +247 \kms$ channel; the background is white in regions for which we have no GBT data. RMs are shown as red ($\RM > 0$) and blue ($\RM < 0$) circles. The magnitude of the RM is proportional to the diameter of the circle, with the largest circles corresponding to $|\RM| \ge 200 \radmsq$. The scale bar assumes a distance of $12.5 \kpc$.
}
\label{fig:hi_wide_map}
\end{figure*}

\begin{figure*}[tb]
\plotone{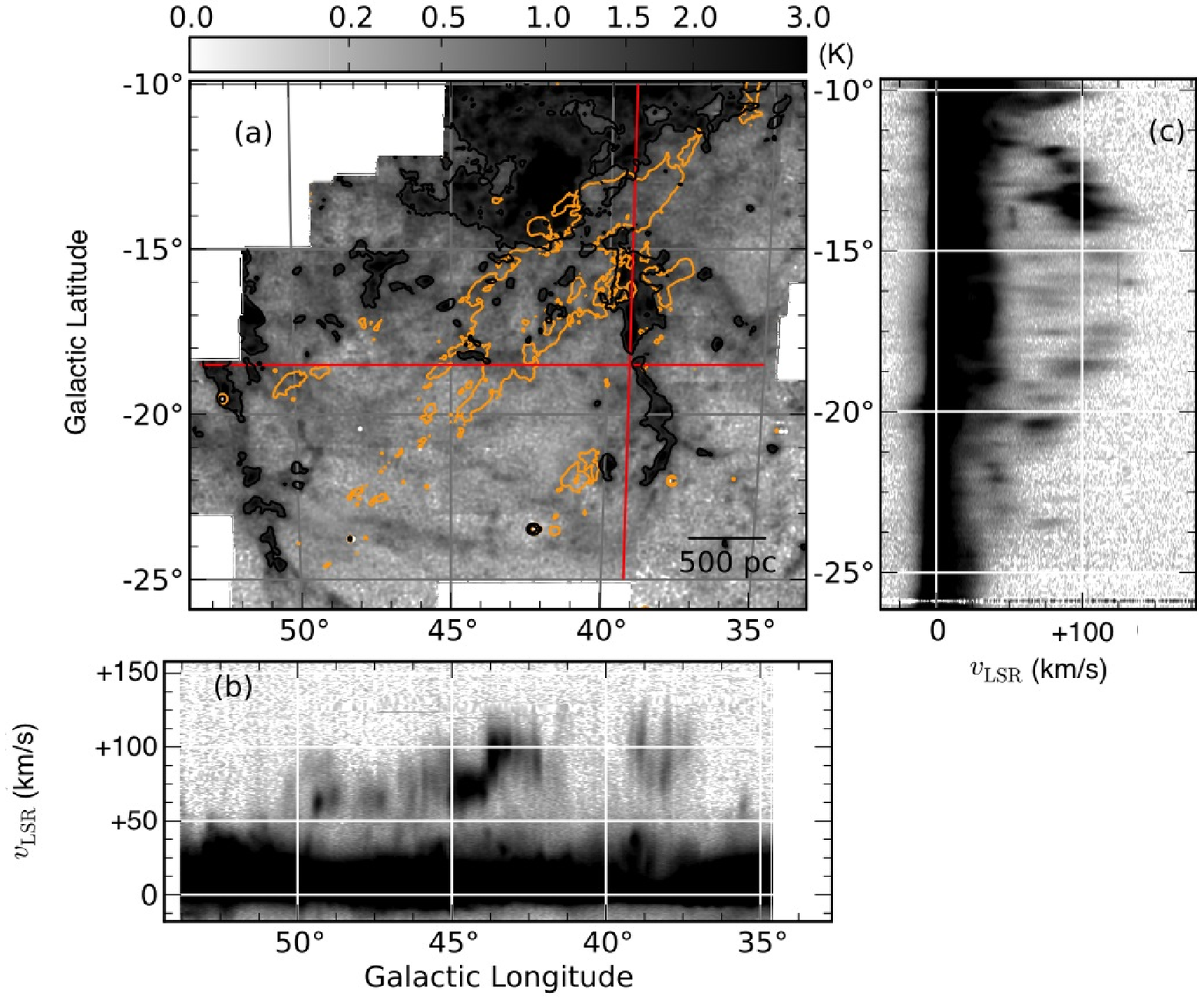}
\caption{Panel $a$ shows a comparison of \ion{H}{1} at the Smith Cloud velocity and at lower velocities. The grayscale shows \ion{H}{1} in the $\vlsr = +40.3 \kms$ channel with a square root intensity scale, with black contours of the same data set at $1.6 \K$. The orange contours show the data from Figure~\ref{fig:hi_wide_map} at $0.5 \K$. The velocity $\vgsr = +247 \kms$ corresponds to $\vlsr = +115 \kms$ at $(l,b) = (38\arcdeg,-13\arcdeg)$ and $\vlsr = +98 \kms$ at $(47\arcdeg, -22\arcdeg)$. Panels $b$ and $c$ show longitude-velocity and latitude-velocity diagrams along the red lines in panel $a$.}
\label{fig:hi_map}
\end{figure*}

\begin{figure*}[tb]
\plotone{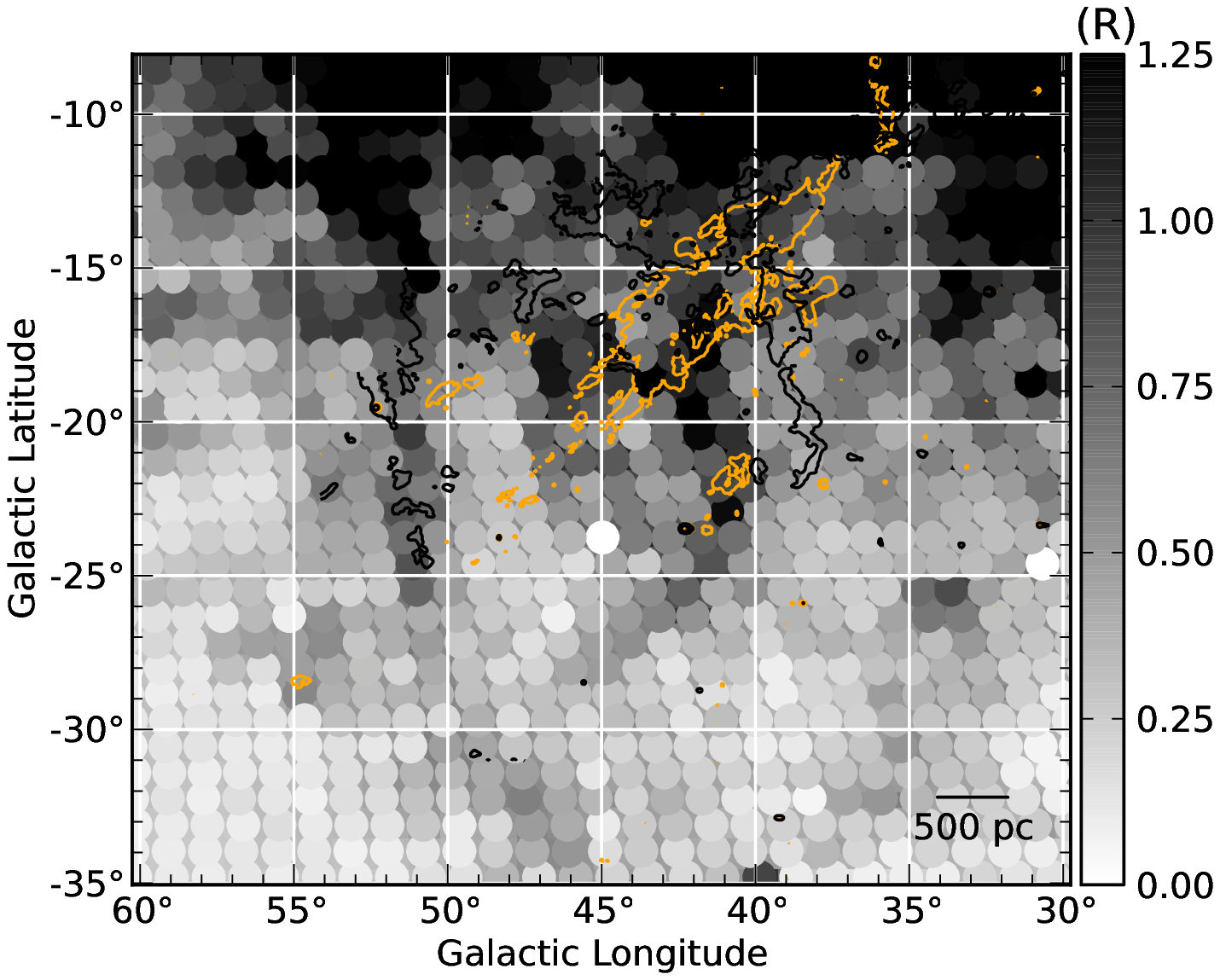}
\caption{Emission from ionized and neutral gas in the Smith Cloud region. The grayscale shows WHAM-NSS data integrated over $+25 \kms < \vlsr < +50 \kms$, with the gray circles denoting WHAM beams. The orange and black contours show \ion{H}{1} data exactly as in Fig.~\ref{fig:hi_map}.}
\label{fig:ha_map}
\end{figure*}

\begin{figure*}[tb]
\plotone{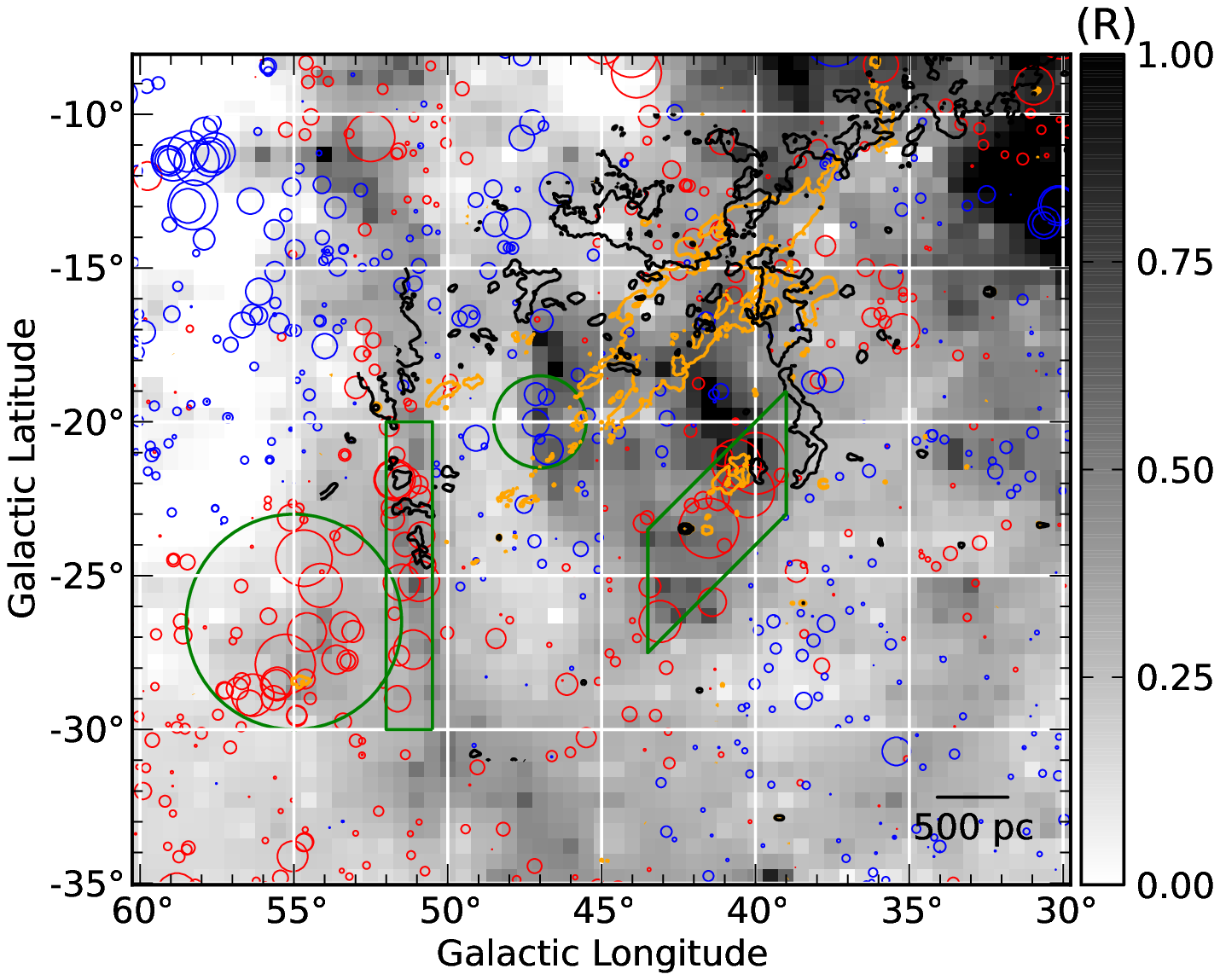}
\caption{Fit-subtracted RMs (\RMHVC) overlaid on \Ha\ and \ion{H}{1} data. The grayscale shows WHAM-NSS \Ha\ data integrated over $+25 \kms < \vlsr < +50 \kms$. We have subtracted the estimated Sagittarius Arm contribution (eq.~\ref{eq:Sag}) and smoothed the \Ha\ observations from the WHAM beams in Fig.~\ref{fig:ha_map}. The \ion{H}{1} contours are as in Fig.~\ref{fig:hi_map}. The red and blue circles denote \RMHVC\ (Section~\ref{sec:rmsubtract}); as in Fig.~\ref{fig:hi_wide_map}, the largest-diameter symbols have $|\RMHVC| = 200 \radmsq$. The green shapes denote regions with positive and negative \RMHVC\ values which we use to calculate $B_{||}$ in Table~\ref{tbl:B}.}
\label{fig:everything_map}
\end{figure*}

\begin{figure}[tb]
\plotone{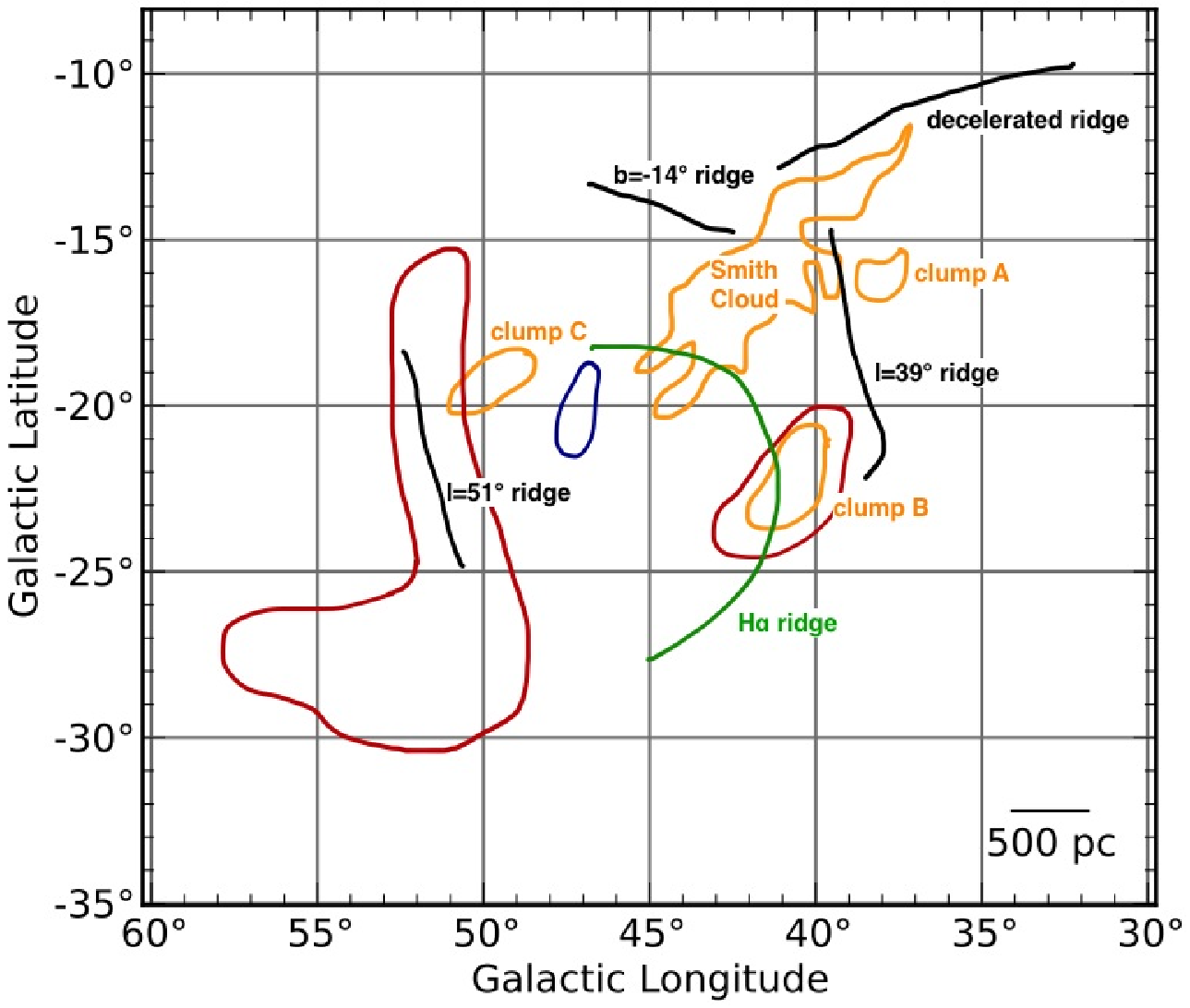}
\caption{Schematic diagram of the features in the Smith Cloud, drawn to scale in comparison to Figs.~\ref{fig:ha_map} and \ref{fig:everything_map}. The $\vgsr \approx +230 \kms$ \ion{H}{1} emission from the Smith Cloud is outlined in orange. Prominent $\vlsr \approx +40 \kms$ \ion{H}{1} features are shown in black, and the $+40 \kms$ \Ha\ ridge in green. Red and blue circular regions denote concentrations of positive and negative RMs (respectively) which are likely associated with the Smith Cloud.}
\label{fig:schematic}
\end{figure}

\section{Description of Smith Cloud environs} \label{sec:overview}

The region around the Smith Cloud is morphologically and kinematically complex. Here, we describe and identify features evident in the data (Figures~\ref{fig:hi_wide_map}$-$\ref{fig:everything_map}) and sketched in Figure~\ref{fig:schematic}. In Section~\ref{sec:interaction}, we interpret the morphology as the result of an interaction between the Smith Cloud and the ambient ISM.

\subsection{Neutral hydrogen emission} \label{sec:hi}

The Smith Cloud shows a gradient in \vlsr\ from $+100 \kms$ at its head to $+70 \kms$ at its tail, which can be understood as the change of the Sun's projected motion over the large angle. In the Galactic standard of rest (GSR) frame (the frame in which the Galactic center is at rest) the hydrogen has a uniform velocity, $\vgsr \sim +230 \kms$. In Figure~\ref{fig:hi_wide_map}, we show the \ion{H}{1} emission at $\vgsr = +247 \kms$. In addition to the main head-tail cloud extending from $(l,b) \approx (38\arcdeg,-13\arcdeg)$ to $(47\arcdeg,-22\arcdeg)$, there are several knots of emission spatially offset from the cloud but at the same \vgsr. One near $(55\arcdeg, -28\arcdeg)$ is close to the major axis of the cloud, but there are also clumps of emission offset from the major axis of the cloud, near $(38\arcdeg, -17\arcdeg)$, $(41\arcdeg, -22\arcdeg)$, and $(49\arcdeg, -18\arcdeg)$. These three clumps are identified as clump A, B, and C, respectively, in Figure~\ref{fig:schematic}.

Lower-velocity \ion{H}{1} emission at $\vlsr \approx +40 \kms$ is also likely associated with the Smith Cloud. The \ion{H}{1} ``decelerated ridge'' along the northwest side of the Smith Cloud was identified as gas decelerated from the Smith Cloud by L08. Figure~\ref{fig:hi_map} shows additional relatively straight ridges in intermediate velocity \ion{H}{1} at $\vlsr \approx +40 \kms$, identified as the ``$l = 39\arcdeg$ ridge'' and ``$b = -14\arcdeg$ ridge'' in Figure~\ref{fig:schematic}. These ridges extend away from near the midpoint of each side at $\approx 45\arcdeg$ angles relative to the major axis of the Smith Cloud.\footnote{Because of the contamination of $\vlsr \approx 0$ local gas, the GSR frame is not useful for identifying the decelerated gas associated with the Smith Cloud.}

The $b=-14\arcdeg$ ridge is somewhat difficult to disentangle from foreground emission due to its lower latitude. Thus, we focus on the $l=39\arcdeg$ ridge, shown in a latitude-velocity diagram in Figure~\ref{fig:hi_map}$c$. The emission extends from $\vlsr = +100 \kms$ at $(39\arcdeg, -16\arcdeg)$, the edge of the main portion of the Smith Cloud, until it merges with the $\vlsr \approx +40 \kms$ emission near $b = -21\arcdeg$. The $\vlsr \approx +40 \kms$ emission in Figure~\ref{fig:hi_map} is anti-correlated with the $\vgsr = +247 \kms$ emission near $(41\arcdeg, -14.5\arcdeg)$, indicating interaction between gas at the two velocities.

\subsection{Ionized hydrogen emission} \label{sec:hplus}

\citet{Hill:2009gx} presented a map of \Ha\ emission at $\vlsr \approx +100 \kms$. The morphology of the \Ha\ emission is smoother than that of the \ion{H}{1}, with a typical intensity of $0.1-0.3 \R$ in nearly all directions which contain $\vlsr \approx 100 \kms$ \ion{H}{1} emission, both along the main axis of the cloud and in clumps A, B, and C. For comparison, the typical integrated \ion{H}{1} intensity in clump B is $\approx 40\%$ of the integrated intensity along the main axis of the Smith Cloud. The relative lack of structure in \Ha\ compared to \ion{H}{1} is probably due in part to the differing angular resolutions of $1\arcdeg$ and $9\arcmin$, respectively.

The \Ha\ emission at $\vlsr \approx +40 \kms$ (Fig.~\ref{fig:ha_map}) is more structured than the $+100 \kms$ emission. To the east and south of the $l=39\arcdeg$ and $b=-14\arcdeg$ \ion{H}{1} ridges is a roughly semicircular enhancement in \Ha\ (Fig.~\ref{fig:ha_map}) called the ``\Ha\ ridge'' in Figure~\ref{fig:schematic}; this ridge is most clear in Figure~\ref{fig:everything_map} (described in Section~\ref{sec:B}). The \Ha\ ridge is separated from the \ion{H}{1} in the $l=39\arcdeg$ ridge by $\approx 2\arcdeg$, and the $l=39\arcdeg$ ridge emission curves eastward outside the \Ha\ ridge near $(l, b) = (38\arcdeg, -21\arcdeg)$. The brightest emission in the \Ha\ ridge is in a filament $\le 1$ WHAM beam across with $I_{\Ha} \approx 0.6-1.2 \R$; fainter emission ($\approx 0.6 \R$) is evident inside the \Ha\ ridge.

\subsection{Faraday rotation}

The raw RMs from \citet{Taylor:2009hv} are shown in Figure~\ref{fig:hi_wide_map}. Off the major axis of the cloud, in the upper left portion of the figure, the RMs are relatively large and negative (shown as blue), while in the lower right portion of the figure, the RMs are relatively small but still negative. The distribution of RMs appears similar in the main portion of the Smith Cloud and outside, suggesting that the $\vgsr \approx +230 \kms$ \ion{H}{1} does not contribute significantly to the RMs. There are positive RMs (shown as red) near the cloud but offset from the brightest $\vgsr = +247 \kms$ emission, with particularly strong concentrations of positive RM near clump B and the \Ha\ ridge, along the $l=51\arcdeg$ ridge, and to higher longitudes, near $(l, b) = (55\arcdeg, -18\arcdeg)$. This emission is coincident with the edge of the fainter \Ha\ emission associated with the $l=51\arcdeg$ ridge.

\label{sec:rmsubtract}
The large-scale distribution of RMs is well fit by a linear surface, which we interpret as a foreground contribution due to the magneto-ionic medium of the Galaxy \citep[following][]{McClureGriffiths:2010fc}. To model the contribution of the foreground to the Faraday rotation, we performed a 2D fit of all the RMs in Figure~\ref{fig:hi_wide_map} as a function of $l$ and $b$. The fit RMs are $-89 \radmsq$ at $(60\arcdeg, -10\arcdeg)$ and $-27 \radmsq$ at $(30\arcdeg, -25\arcdeg)$, consistent with the wide-area fits by \citet{Taylor:2009hv}. We subtracted this fit from the observed RMs; the difference, \RMHVC, represents magnetized ionized gas associated with the Smith Cloud. We show \RMHVC\ in Figure~\ref{fig:everything_map}. There is no clear evidence of non-zero \RMHVC\ in the main portion of the Smith Cloud, where the \ion{H}{1} at $\vgsr \sim +230 \kms$ is brightest.

Figure~\ref{fig:everything_map} shows that the regions of enhanced RM are coincident with enhancements in $\vlsr \approx +40 \kms$ \Ha\ emission. Two regions of enhanced Faraday rotation are coincident with the \Ha\ ridge, with $\RMHVC \approx +120$ to $+250 \radmsq$ near $(l,b) = (41\arcdeg, -22\arcdeg)$ and $\RMHVC \approx -50$ to $-100 \radmsq$ near $(47\arcdeg, -20\arcdeg)$. The RM enhancements are each coincident with \Ha\ emission slightly fainter than the brightest emission in the \Ha\ ridge. We note that the enhancement in negative RMs near $(58\arcdeg, -11\arcdeg)$ is in a region with bright $\vlsr \approx 0$ \Ha\ emission which extends off the figure towards the plane and Galactic east, so this RM enhancement is likely associated with a foreground \ion{H}{2} region.

\section{Interaction with the halo ISM} \label{sec:interaction}

The morphology and kinematic structure of the features described in the previous section suggest that the gas responsible for the $\vgsr \approx +230 \kms$ (or $\vlsr \approx +100 \kms$ near the head of the cloud) and the $\vlsr \approx +40 \kms$ \ion{H}{1} and \Ha\ emission are physically associated and likely the result of interaction with the ambient ISM (L08). This is most clearly indicated by the anti-correlation of the $\vlsr \approx +40 \kms$ \ion{H}{1} emission from the $l=39\arcdeg$ ridge (black contours in Fig.~\ref{fig:hi_map}) and the $\vgsr \approx +230 \kms$ (orange contours). The connection between the $\vlsr = +40$ and $+100 \kms$ emission along the $l=39\arcdeg$ ridge is also evident in Figure~\ref{fig:hi_map}$c$. The similar morphology of the \Ha\ ridge and the \ion{H}{1} $l=39\arcdeg$ ridge, both at $\vlsr \approx +40 \kms$, strongly suggests that they are, in turn, physically associated. The dominance of \Ha\ in some regions and \ion{H}{1} in others may indicate shock ionization in some regions of the cloud or may reflect an emission measure or column density threshold above which neutral gas is shielded from photoionization by the Galactic radiation field.

The \ion{H}{1} mass in the decelerated ridge is too large to account for with swept-up ISM gas that far from the Galactic plane. L08 thus argued that the decelerated ridge consists of material which has been stripped from the Smith Cloud and decelerated by the interaction with disk gas. The same argument applies to the $b=-14\arcdeg$ ridge (which may be the same structure as the decelerated ridge) and the $l=39\arcdeg$ ridge;
the latter has an \ion{H}{1} mass of $\sim 5 \times 10^5 M_\odot$, assuming the gas is at the Smith Cloud distance.

The $l = 51\arcdeg$ ridge is connected in velocity space to the main emission from the cloud, as shown in the longitude-velocity diagram in Figure~\ref{fig:hi_map}$b$, suggesting physical association. The emission from the ridge at $l \approx 50\arcdeg$, $\vlsr = +40 \kms$ connects to the tail of the \ion{H}{1} emission associated with the main Smith Cloud ($l=44\arcdeg$, $\vlsr=+80 \kms$; see the orange contours in Figure~\ref{fig:hi_map}). However, in contrast to the bow wave-like appearance of the $l=39\arcdeg$ and $b=-14\arcdeg$ ridges with respect to the Smith Cloud, the \ion{H}{1} and \Ha\ morphology of the $l=51\arcdeg$ ridge does not clearly indicate interaction with the $\vgsr \approx +230 \kms$ \ion{H}{1}. Therefore, it is possible that this ridge is a foreground feature.

Because the majority of the $\vlsr \sim +40 \kms$ emission in both \ion{H}{1} and \Ha\ is associated with the $\vgsr \approx +230 \kms$ emission, we adopt the Smith Cloud distance of $12.5 \pm 1.3 \kpc$ \citep[][L08]{Wakker:2008fo} for all of this emission for the remainder of this paper.

\section{Electron density and magnetic field strength} \label{sec:B}

The correlation of enhancements in RM with decelerated \Ha\ structures associated with the Smith Cloud suggests that \RMHVC\ is due to magnetic fields in the Smith Cloud. To estimate the magnetic field strength, we first estimate the electron density from the \Ha\ emission. To derive the \Ha\ emission at $\vlsr \approx +40 \kms$ due to the Smith Cloud, we estimate the \Ha\ contribution of the warm ionized medium in the foreground Sagittarius Arm assuming \citep{Haffner:1999hi}
\begin{equation} \label{eq:Sag}
I_\mathrm{Sgr}(b) = I_0 \exp \left(- \frac{2D}{H} \tan |b|\right).
\end{equation}
We performed a weighted linear fit of the \Ha\ emission for latitudes $-0.1 > \tan b > -0.7$ integrated from the WHAM-NSS data over the velocity range $+25 \kms < \vlsr < +50 \kms$ as a function of $\tan b$. We used the median value over the longitude range $30\arcdeg < l < 60\arcdeg$ following the procedure \citet{Haffner:1999hi} used for the Perseus Arm. This \Ha\ emission is well fit by equation~(\ref{eq:Sag}) with the fit parameters $\ln (I_0/\R) = 0.93 \pm 0.12$ and $2D/H = 4.38 \pm 0.35$. We used the mean absolute deviation about the median as the uncertainty in $I_{\mathrm{Sgr}}$. Here $D$ is the distance to the emission; the emission is likely distributed along distances from $\sim 2-10 \kpc$ because the arm is near tangency at this longitude \citep{Benjamin:2008uh}. $D$ is also double-valued \citep{Gomez:2006bn}. The warm ionized medium scale height is $H$. Because $H \approx 1 \kpc$ for the ionized gas but $H \approx 300 \pc$ for the neutral gas \citep[e.g.][]{Cox:2005vl}, the foreground contribution of the Sagittarius Arm to the \ion{H}{1} at these latitudes is negligible.

We derive the observed emission measure due to the Smith Cloud from the \Ha\ intensity in each beam,
\begin{eqnarray} \label{eq:EM}
\EM &\equiv& \int n_e^2(s) ds \nonumber \\
&=& 2.75 \left(\frac{T}{10^4 \K}\right)^{0.9} \left(\frac{I_{\Ha} - I_\mathrm{Sgr}(b)}{\textrm{R}}\right) \pc \cmsix,
\end{eqnarray}
assuming $T = 8000 \K$ in the \Ha-emitting gas. Because the underlying density distribution $n_e(s)$ is uncertain, we assume the simplest case: $n_e(s)$ is a step function with a value of $n_e$ or zero along the line of sight, and the path length in which $n_e(s) = n_e$ is $L_{\Hp} \equiv fL$, where the gas occupies a fraction $f$ of the volume along the total path length $L$. Then the electron density is
\begin{equation} \label{eq:ne}
n_e = \sqrt{\EM / L_{\Hp}}.
\end{equation}
If the gas density is determined by a series of compressions and rarefactions such as might be produced by turbulence, we would expect a lognormal distribution of density \citep{VazquezSemadeni:1994fm}. As a test, we replaced the uniform distribution assumed above with a lognormal distribution of electron density with a width appropriate for the warm ionized medium \citep{Hill:2008gv}. This yields only a $\sim 0.9$ correction to the derived most probable density, so we retain the uniform assumption for simplicity.


\begin{deluxetable}{rrr r@{$\, \pm \,$}l r@{$\, \pm \,$}l r r@{$\, \pm \,$}l}
\tabletypesize{\footnotesize}
\tablecolumns{10}
\tablewidth{0pt}
\tablecaption{Magnetic field estimates\label{tbl:B}}
\tablehead{\colhead{$l$} & \colhead{$b$} & \colhead{$N$} & \mc{$\langle \RMHVC \rangle$} & \mc{$\langle \EM \rangle$} & \colhead{$L_{\Hp}$} & \mc{$\langle B_{||} \rangle$} \\
 & & & \mc{($\textrm{rad m}^{-2}$)} & \mc{($\textrm{pc cm}^{-6}$)}	& \colhead{(pc)}	& \mc{($\mu \mathrm{G}$)} }
\startdata
$41\arcdeg$ & $-22\arcdeg$ & $17$ & $+108$&$ 3$ & $1.22$&$0.04$ & $ 220$ & $+8.2$&$0.3$ \\
$46\arcdeg$ & $-19\arcdeg$ & $ 7$ & $-51$&$ 4$ & $1.00$&$0.08$ & $ 220$ & $-4.3$&$0.4$ \\
$51\arcdeg$ & $-25\arcdeg$ & $16$ & $+90$&$ 3$ & $0.76$&$0.04$ & $ 220$ & $+8.6$&$0.3$ \\
$54\arcdeg$ & $-26\arcdeg$ & $30$ & $+73$&$ 1$ & $0.38$&$0.03$ & $1000$ & $+4.6$&$0.2$ 
\enddata
\tablecomments{\RMt\ and $\langle \EM \rangle$ are measured within the green shapes in Fig.~\ref{fig:everything_map}. $L_{\Hp}$ is an assumed upper limit based on the \Ha\ morphology. $|B_{||}|$ is the lower limit derived from equation~(\ref{eq:B}). The number of WHAM beams in the region is $N$.}
\end{deluxetable}

The line-of-sight component of the magnetic field is
\begin{equation} \label{eq:B}
\frac{\langle B_{||} \rangle}{\uG} = \frac{\RMt}{0.81 \, \langle n_e \rangle \, L_{\Hp}}
= \frac{\RMt}{0.81 \times (L_{\Hp} \langle \EM \rangle)^{1/2}},
\end{equation}
under the additional assumption that the field does not vary along the line of sight with \RMt, $n_e$, and $L_{\Hp}$ in\radmsq, $\mathrm{cm}^{-3}$, and pc, respectively. Here \RMt\ is the weighted mean of \RMHVC\ for all sources in a defined region. The weights are $w_i = \sigma_i^{-2} / ( \Sigma_i \sigma_i^{-2} )$, where $\sigma_i$ is the quadrature sum of the statistical uncertainty in ${\RMHVC}_{,i}$ and $7 \radmsq$, an estimate of the intrinsic variation in the extragalactic radio sources \citep{Schnitzeler:2010in,Stil:2011gf}. There is an additional, systematic uncertainty due to our Milky Way foreground RM subtraction which is not included in our uncertainty. The uncertainties reported in Table~\ref{tbl:B} are the standard deviation of the weighted mean, $(\Sigma_i w_i^2 \sigma_i^2)^{1/2}$. 

We determine the average field for four regions. Our results are shown in Table~\ref{tbl:B}. 
We choose $L_{\Hp}$ to be the largest value which is reasonable based upon the morphology, so our estimates of the magnetic field are lower limits. Along the \Ha\ ridge (first two rows of Table~\ref{tbl:B}), the brightest \Ha\ emission and the RM enhancement are one WHAM beam wide, so we assume that $L_{\Hp} < 220 \pc$, the projected beam size. The $l=51\arcdeg$ ridge (third row of Table~\ref{tbl:B}) is also $\lesssim 1$ WHAM beam wide; if this ridge is a foreground feature (see above), $L_{\Hp}$ would be smaller and $B_{||}$ larger than we estimate. The downstream ($53\arcdeg \lesssim l \lesssim 57\arcdeg$) region of enhanced \Ha\ and \RMHVC\ is larger; we estimate $L_{\Hp} \lesssim 1000 \pc$. 

%

\section{Discussion} \label{sec:discussion}

The morphologies of the \Ha\ and \ion{H}{1} emission at different velocities and of the RMs provide insight into the physical processes at work.

\subsection{Shock ionization or photoionization?}

The \Ha\ emission at $\vlsr \approx +100 \kms$ is relatively smooth, with the exception of the tip near $(39\arcdeg,-13\arcdeg)$, with an observed \Ha\ intensity of $\approx 0.1-0.3 \R$ \citep{BlandHawthorn:1998kb,Hill:2009gx}. The smooth emission suggests photoionization, consistent with the accurate prediction of the \Ha\ intensity from the Smith Cloud by a model of the Galactic ionizing radiation field \citep{Putman:2003ba}. In this scenario, the gas from the cloud would be largely ionized in a skin up to a constant emission measure as seen from the midplane, $\EM_{\perp}$, while gas beyond this threshold would be neutral. Assuming that the depth of the ionized skin, $L_{\perp} = \EM_{\perp} n_e^{-2}$, is small enough so that $L_{\perp} / \tan |b|$ is less than the width of the cloud projected onto the plane, we have $\EM_{\perp} = \EM \sin |b|$. From the observed \Ha\ intensities, $\EM_{\perp} \approx 0.05 - 0.2 \pc \cmsix$ (varying based upon the assumed extinction correction and temperature). This photoionization scenario explains the \Ha\ emission from most other HVCs in which \Ha\ has been detected \citep{Tufte:1998cb,Tufte:2002du,Haffner:2005td,Putman:2003ba,Barger:2012jz}.

The intensity of the $\vlsr \approx +40 \kms$ emission from the \Ha\ ridge is less smooth, has a higher \Ha\ intensity of $\approx 0.4-0.7 \R$ (Table~\ref{tbl:B}), and does not have coincident \ion{H}{1}, so it is more difficult to explain this emission as photoionization by the Galactic radiation field. Ionization related to shocks may better explain this emission, either directly or through photoionization from shock emission \citep{BlandHawthorn:2007kc}. Given the sound speed in $8000 \K$ ionized gas of $10 \kms$ and the velocity of the Smith Cloud with respect to the corotating ISM of $130 \kms$ (L08), shocks are expected, although the ambient density is uncertain and presumably small.

The velocity is sufficient to produce some \ion{Si}{4}, \ion{C}{4}, \ion{N}{5}, and \ion{O}{6}, with the ratios depending upon the fraction of the kinetic energy which converts to magnetic rather than thermal energy \citep{Allen:2008gj,Henley:2012hy}. The shock is slow enough so that the majority of this emission would be in the shock, not the post-shock cooling-zone, so the expected shock is in a regime in which the ratios of these lines are useful diagnostics of the shock conditions.  None of the sightlines probed with the {\em Far Ultraviolet Spectroscopic Explorer} by \citet{Sembach:2003dr} or others are near the Smith Cloud. The \ion{Si}{4}, \ion{C}{4}, and \ion{N}{5} lines are all accessible with the Cosmic Origins Spectrograph on the {\em Hubble Space Telescope}. Optical forbidden line emission, accessible with WHAM, can also constrain shock ionization scenarios. \citet{Hill:2009gx} did not detect \oiii $\lambda 5007$ from the tip of the cloud, although \citet{Allen:2008gj} only predict strong \oiii\ in a limited range of shock velocities around $100-200 \kms$.

\label{sec:shocks}

\citet{Hill:2009gx} measured an optical line ratio $\nii/\Ha = 0.32 \pm 0.05$ towards the tip of the Smith Cloud, which they used to estimate a nitrogen abundance $\mathrm{N}/\mathrm{H} = 0.15-0.44$ times solar. This estimate was based on the assumption that the gas is photoionized. However, their measurement was towards the portion of the cloud at the leading edge, the brightest portion of the HVC in both \Ha\ and \ion{H}{1} at $\vlsr \approx +100 \kms$. Both the $+100 \kms$ \Ha\ intensity (0.43~R) and the geometry suggest that shocks may be located within this beam, potentially making the photoionization assumption invalid. The \sii\ line width in this beam implies a nonthermal velocity of $11.5 \pm 3.4 \kms$ \citep{Hill:2009gx}. \citet{Putman:2003ba} measured $\nii / \Ha = 0.6$ in a downstream portion of the main body of the cloud; in the \Ha\ map of \citet{Hill:2009gx}, this region has smoother emission more suggestive of photoionization. With an assumed temperature of $8000-12000 \K$, typical of low-density photoionized gas \citep{Madsen:2006fw}, the nitrogen abundance implied by $\nii / \Ha = 0.6$ from Figure~4 of \citet{Hill:2009gx} is $0.3 < (\mathrm{N}/\mathrm{H}) / (\mathrm{N}/\mathrm{H})_{\odot} < 0.8$. We note that the shocked region at the tip of the cloud may be quite small; it is not sampled by any RMs used in the present paper.

\subsection{Magnetic field geometry}

The enhanced Faraday rotation is better correlated with filaments of $\vlsr \approx +40 \kms$ \Ha\ emission than any other emission tracer. 
To constrain the cause, we estimate the electron column density, $n_e L_{\Hp}$.
For the observed emission measures and assuming a constant electron density, the column density of $\vlsr \approx +40 \kms$ \Hp\ is
$
n_e L_{\Hp} \sim (\EM \, L_{\Hp})^{1/2} = 4 \times 10^{19} \cm^{-2} (L_{\Hp} / 200 \pc)^{1/2}
$
in the \Ha\ ridge,\footnote{This corresponds to a mass integrated over $+25 \kms < \vlsr < +70 \kms$ in the $80$ beams within $5\arcdeg$ of $(43.5\arcdeg, -21\arcdeg)$ for which $\EM > \sigma_{\EM}$ of $M_{\Hp} \approx 1 \times 10^6 (L_{\Hp} / 200 \pc)^{1/2} M_{\odot}$.} compared to $(1 - 3) \times 10^{19} \cm^{-2}$ for the diffuse, $\vlsr \approx +100 \kms$ \Ha\ \citep{Hill:2009gx}. Therefore, if $L_{\Hp} \sim 200 \pc$ (our best estimate; see Section~\ref{sec:B}) in the ridges which contribute most to the observed RMs, $n_e \, L_{\Hp}$ is similar in the ridges and in the diffuse \Ha-emitting gas. We would then expect a comparable $\RM \sim B_{||} n_e L_{\Hp}$ across the cloud to that in the filament if $B_{||}$ were constant.

Instead, the RMs through the bulk of the \ion{H}{1} emission from the Smith Cloud (Figures~\ref{fig:hi_wide_map} and \ref{fig:schematic}) are consistent with the foreground. This could be explained by a larger $B_{||}$ in the \Ha\ ridge or by a magnetic field which is less ordered in the Smith Cloud than in the \Ha\ ridge, producing more field reversals along the line of sight. Either scenario could be explained by a physical compression of the gas, which would draw together and order field lines.


In addition to the sightlines with $|\RMHVC| \gtrsim 100 \radmsq$ which are correlated with filaments, the RMs downstream of the \ion{H}{1} and \Ha\ emission from the Smith Cloud (in the lower left quadrant of Fig.~\ref{fig:everything_map}) are positive compared to the foreground. This may suggest an extended, low column density wake which is ionized and magnetized but at too low a column to detect in emission tracers. However, the association of this downstream gas with the Smith Cloud is less certain than for the \Ha\ ridge and the $l=39\arcdeg$ and $b=-14\arcdeg$ ridges.

Faraday rotation is sensitive only to magnetic fields in ionized gas. Although magnetized gas in the Galaxy with a low ionization fraction does not contribute significantly to Faraday rotation, gas with the ionization fraction of the warm and cold neutral media is still affected by Lorentz forces due to ion-neutral collisions \citep{Kulkarni:1987th,Ferriere:2001vr}. Therefore, if a field is present in the neutral gas but does not have an evident RM signature due to the low $n_e$, it could still be dynamically important. Because most of the RM detections reported here are concentrated in narrow, possibly-shocked filaments, the magnetic field lines are likely compressed. Because the mass contained in the filaments is larger than can be explained by swept-up gas from the ambient ISM, the filaments are most likely gas stripped and decelerated from the HVC.

The positive values of \RMHVC\ on the $l \approx 41\arcdeg$ side of the \Ha\ ridge and the negative values of \RMHVC\ on the $l=47\arcdeg$ side of the \Ha\ ridge (see Figure~\ref{fig:everything_map}), combined with the near-zero \RMHVC\ values along the ridge between these data, suggest a toroidal field pointing towards the observer at $l \approx 41\arcdeg$, away from the observer at $l \approx 47\arcdeg$, and perpendicular to the line of sight in between. In 2D magnetohydrodynamic (MHD) simulations, \citet{Konz:2002cy} found that an HVC moving through a magnetized medium would establish a magnetic barrier which provides thermal insulation between the cool cloud and the hot ambient medium, reducing the disruption of the cloud. A magnetic field should also reduce Kelvin-Helmholtz instabilities in the interface between the cloud and the ambient medium. In the \citet{Konz:2002cy} model, an HVC can compress field lines in a much weaker ambient field to create a field of a few \uG\ in the head of a cloud, qualitatively similar to the observations we report here.

The peak magnetic field of $\gtrsim 8 \uG$ in the Smith Cloud is stronger than the typical field in the Galactic midplane of a few \uG, particularly the regular (non-turbulent) component of $\sim 1-2 \uG$ \citep{Ferriere:2001vr}. Estimates of the field at $|z| \approx 3 \kpc$ at a Galactocentric radius of $\approx 8 \kpc$ based on comparisons of models to observed RMs or synchrotron emissivity are all $\lesssim 2.5 \uG$ and typically $\approx 1 \uG$ \citep{Cox:2005vl,Mao:2010eg,Mao:2012kq,Jansson:2012ep}. However, these measurements are all made locally and vary between magnetic spiral arms \citep{Jansson:2012ep}.
These weak Milky Way halo magnetic field strength measurements in an environment similar to that surrounding the Smith Cloud support our interpretation that the ambient field must have been compressed in order to produce the strong observed field. 


\subsection{Implications and future prospects}

The observed morphology of the neutral and ionized gas and the Faraday rotation described here promise to provide an excellent benchmark for 3D MHD simulations of HVCs entering the Galactic disk. For example, such simulations could help to determine whether the observed RM distribution is due primarily to stronger fields in the decelerated gas or to observational selection effects related to either the orientation of the field or the electron density. MHD simulations could also test whether small seed fields in the HVC could be amplified to produce the observed fields, since existing models rely upon the swept-up ambient field. Such models will improve our understanding of the role the magnetic field plays in regulating the disk-halo interaction.

The technique for measuring magnetic fields in isolated interstellar clouds developed by \citet{McClureGriffiths:2010fc} and expanded here can be used in any region in which \Ha\ spectra (or some other estimate of the electron column) and background RMs are available. The \citet{Taylor:2009hv} catalog of RMs covers the northern sky ($\delta > -40\arcdeg$) with $\approx 1$ source deg$^{-2}$ on average. The technique works best in regions of the sky with relatively smooth large-scale features in Faraday rotation, as RMs from background sources do not provide depth information to separate out foregrounds. WHAM-NSS provides \Ha\ spectra for $\delta > -30\arcdeg$, and the observations for the WHAM Southern Sky Survey, which will fill in the remainder of the sky, are complete and now being reduced \citep[Hill et al in prep]{Haffner:2010tha}. WHAM survey data only provide complete \Ha\ velocity coverage to $|\vlsr| < 80 \kms$, but WHAM is also used to obtain spectra of HVCs in other velocity windows \citep{Tufte:1998cb,Tufte:2002du,Haffner:2005td,Hill:2009gx,McClureGriffiths:2010fc,Barger:2012jz}.

The new broadband backends on both the Karl G.\ Jansky Very Large Array \citep{Perley:2009fp} and the Australia Telescope Compact Array \citep{Wilson:2011bg} make these telescopes efficient machines for measuring Faraday rotation towards point sources, making RMs accessible with short exposure times over the full sky for much fainter sources than \citet{Taylor:2009hv} include.
Future telescopes with phased array feeds like the Australian Square Kilometre Array Pathfinder \citep[ASKAP;][]{Johnston:2008gq} and the Square Kilometre Array will provide far more RM measurements over the entire southern sky. In particular, the ASKAP Polarization Sky Survey of the Universe's Magnetism \citep[POSSUM;][]{Gaensler:2010vt} is designed to provide a large, all-southern sky database of RMs which will be well-suited to this kind of work. 

\section{Summary} \label{sec:summary}

Through an analysis of Faraday rotation and spectroscopic maps of \ion{H}{1} and \Ha, we have analyzed the structure of the neutral and magnetoionized gas in the Smith Cloud at a range of velocities. The Smith Cloud is a laboratory in which to study the behavior of a gaseous cloud as it interacts with the high-altitude Galactic ISM. We found that neutral hydrogen which has been stripped and decelerated from the cloud (Section~\ref{sec:hi}; Figure~\ref{fig:hi_map}) has a counterpart a few hundred pc downstream in ionized gas (Section~\ref{sec:hplus}; Figure~\ref{fig:ha_map}) which is most likely shocked. An enhancement in Faraday rotation, relative to the smooth background, is coincident with the decelerated ionized gas (Section~\ref{sec:rmsubtract}; Figure~\ref{fig:everything_map}). Using these observations, we measured a magnetic field in the cloud with a line-of-sight component $\gtrsim 8 \uG$ (Section~\ref{sec:B}). We interpret our results as a signature of compressed and likely shocked gas as it interacts with the ambient ISM. This HVC is an excellent target for future observations of shock diagnostics and MHD simulations to further understand the role magnetic fields play in regulating the accretion of gas into galactic disks.

\acknowledgements

We acknowledge useful discussions with A. J. Fox, B. M. Gaensler, J. A. Green, V. A. Moss, and E. K. Braden. The National Radio Astronomy Observatory is a facility of the National Science Foundation operated under a cooperative agreement by Associated Universities, Inc. The Wisconsin H-Alpha Mapper is supported by the National Science Foundation.

{\it Facilities:} \facility{GBT}, \facility{VLA}, \facility{WHAM}

\bibliography{papers_bibtex}

\end{document}